\documentclass[sigconf]{acmart}

\copyrightyear{2026}
\acmYear{2026}
\setcopyright{cc}
\setcctype{by}
\acmConference[CIKM '26]{Proceedings of the 35th ACM International Conference on Information and Knowledge Management}{November 07--11, 2026}{Rome, Italy}
\acmBooktitle{Proceedings of the 35th ACM International Conference on Information and Knowledge Management (CIKM '26), November 07--11, 2026, Rome, Italy}
\acmISBN{979-8-4007-2539-5/2026/11}
\acmDOI{10.1145/3799682.3840174}
\usepackage{booktabs}
\usepackage{graphicx}
\usepackage{tabularx}
\usepackage{array}
\usepackage{multirow}
\usepackage{xspace}

\newcolumntype{Y}{>{\raggedright\arraybackslash}X}
\newcommand{\tool}{SIDInspector\xspace}
\newcommand{\sid}{SID\xspace}

\title{SIDInspector: A Mapping-First Diagnostic Resource for Semantic-ID Tokenizers}

\author{Jiandong Ding}
\authornote{Corresponding author.}
\email{dingjiandong2@huawei.com}
\affiliation{%
  \institution{Huawei Technologies Co., Ltd.}
  \city{Shanghai}
  \country{China}
}

\author{Heng Chang}
\email{heng.chang@huawei.com}
\affiliation{%
  \institution{Huawei Technologies Co., Ltd.}
  \city{Beijing}
  \country{China}
}

\author{Huijie Qin}
\email{qinhuijie@huawei.com}
\affiliation{%
  \institution{Huawei Technologies Co., Ltd.}
  \city{Shanghai}
  \country{China}
}

\author{Tianying Liu}
\email{liutianying2@huawei.com}
\affiliation{%
  \institution{Huawei Technologies Co., Ltd.}
  \city{Shanghai}
  \country{China}
}

\renewcommand{\shortauthors}{Ding et al.}

\begin{document}

\ccsdesc[500]{Information systems~Recommender systems}
\ccsdesc[300]{Information systems~Retrieval models and ranking}
\keywords{semantic IDs, generative recommendation, tokenizer artifacts, diagnostic evaluation, recommender systems}

\begin{abstract}
Semantic-ID (\sid) tokenizers are increasingly reused as standalone artifacts
in generative recommendation. Once exported, an item-to-code mapping defines
the address space used by a later sequence generator. Yet tokenizer releases
rarely share an inspection interface, so coverage gaps, full-code aliasing,
behaviorally weak prefixes, tail compression, and prefix fan-out may surface
only after downstream training. We introduce \tool, a mapping-first
diagnostic resource for \sid tokenizer artifacts. \tool defines a small adapter
contract over item mappings, metadata, interactions, and optional generator
traces; validates the contract; and reports mapping-level probes for
utilization, aliasing, neighborhood alignment, popularity allocation, and
structural cost, with hooks for temporal churn and generator traces.

We evaluate \tool on four tokenizer lines with explicit provenance. A same-item
Musical study compares controlled GRID-style RQ-KMeans exports at two capacity
settings with a bounded ReSID/GAOQ export; released LETTER and LC-Rec mappings
test the contract on official item-index artifacts. Increasing the GRID-style
budget reduces D2 full-code aliasing from 0.977 to 0.779, while ReSID/GAOQ
reaches 0.000. D3 neighborhood alignment, however, remains below a
deterministic category-prefix control (0.055--0.154 versus 0.447). These
results separate capacity, addressability, and behavioral alignment at the
artifact level. Additional datasets show that D3 scale is dataset-dependent.
A bounded fixed-reranker probe connects D3 to candidate exposure, but final
ranking quality remains a downstream model question.
\end{abstract}

\maketitle
\section{Introduction}

Semantic-ID (\sid) tokenization has become a practical artifact boundary for
generative recommendation. Systems such as TIGER map items into discrete code
sequences so that a sequence model can generate candidate identifiers rather
than score every catalog item directly~\cite{rajput2023tiger}. Once exported,
that item-to-code mapping becomes the address space exposed to the generator.
Recent work changes this address space through residual and recommender-native
quantization~\cite{ju2025grid,liang2026resid}, collaborative or differentiable
tokenization~\cite{zhu2024cost,fu2026diger}, and non-uniform or variable-length
capacity~\cite{wei2026card,cheng2026capsid}. Collision-aware evaluation and
drift-aware refreshes add further artifact-level concerns~\cite{hu2026quasid,zhang2026sidreliability,feng2026dact}.
\tool turns those exports into a shared artifact interface, making
item-to-code mappings inspectable under one contract before downstream
generator training~\cite{chen2026hpgr}.

Reusable tooling has not caught up with this artifact boundary. A new
tokenizer repository may expose checkpoints, item mappings, codebooks, or only
intermediate feature files. Downstream users then face two sets of questions
before training a generator. First, do all catalog items have valid codes, and
how many share a full code? Second, do prefixes reflect user co-occurrence, is
address capacity distributed fairly across the tail, and does the prefix
structure create large fan-out?
Today these checks are usually reimplemented ad hoc inside each paper. The
resulting evidence is difficult to compare because each method reports a
different mixture of codebook usage, downstream ranking, and qualitative
examples.

Aggregate ranking metrics remain necessary, but they do not expose the exported
address space itself. Prior recommender tooling has argued for behavioral tests
beyond NDCG-style aggregates~\cite{chia2022reclist}; \sid tokenizers add a
distinct resource problem: the item mapping should be validated and
profiled before a full generator consumes GPU time. Underused codebooks,
repeated full codes, behaviorally weak prefixes, tail compression, and large
prefix fan-out can be hidden by a downstream score, yet they are direct
properties of the reusable tokenizer artifact.

We present \tool, a diagnostic resource for inspecting item-to-\sid{}
tokenizer artifacts before or alongside downstream recommendation evaluation.
Given exported item codes, it profiles capacity use, aliasing, neighborhood
alignment, head-tail allocation, and structural cost before a full generator
is retrained.

\tool combines three resource components. First, it defines a small adapter
contract for \sid assignments, item metadata, interaction histories, and
optional generator outputs, plus validators that reject incomplete or
ambiguous artifacts before metrics are reported. Second, it implements
D1--D5 mapping-level probes for utilization, aliasing, neighborhood alignment,
popularity allocation, and structural cost, with D6 churn support for
continual-tokenizer settings. Third, worked examples cover four tokenizer
lines. They include GRID/RQ-KMeans-style under two capacity settings and
ReSID/GAOQ on the same Musical items, plus released LETTER and LC-Rec item-index
adapters.
Controls and mechanism probes calibrate the diagnostics. The examples separate
addressability from behavioral neighborhood alignment, two properties that a
method-level score alone does not identify.

\section{Artifact Interface and Workflow}

\begin{figure}[t]
\centering
\includegraphics[width=\columnwidth]{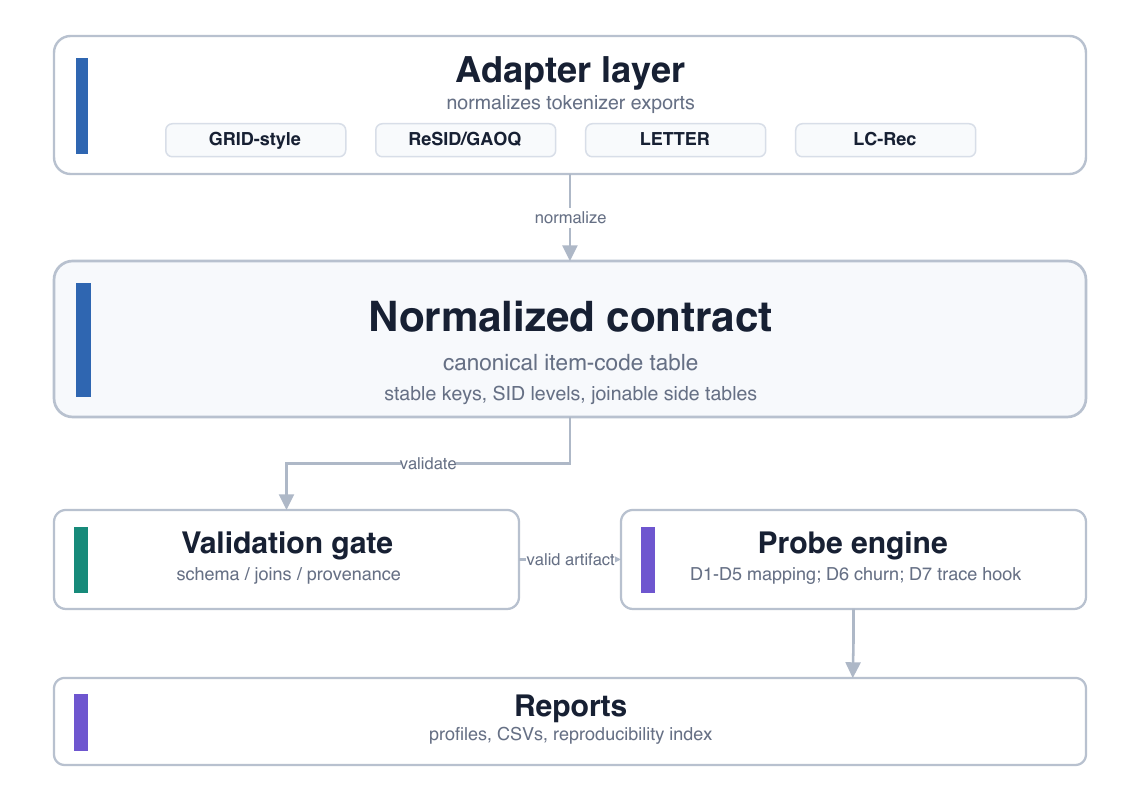}
\caption{\tool workflow. Adapters normalize four V0 artifact lines to a common
item-to-code contract. Validation precedes D1--D5 reports; paired mappings
enable D6 churn, while D7 remains a trace-input hook.}
\Description{SIDInspector resource pipeline. GRID-style, ReSID/GAOQ, LETTER,
and LC-Rec exports pass through the adapter layer into a normalized item-to-code
contract. The validation gate checks schema, joins, and provenance before the
probe engine runs D1 through D5, optional D6 churn, or the D7 trace hook. The
pipeline emits artifact profiles, CSV files, and a reproducibility index.}
\label{fig:audit-sid-pipeline}
\end{figure}

\tool treats a tokenizer export as an inspectable resource artifact, not as an
opaque component of a trained recommender. The minimum input is a table
\texttt{sid\_assignments(item\_id, sid\_0, ..., sid\_L)}. Optional tables provide
metadata, interactions, refresh pairs, and generator outputs. This
mapping-first contract lets adapters normalize heterogeneous
repositories while keeping the inspection independent of a particular training loop.
Figure~\ref{fig:audit-sid-pipeline} summarizes the architecture and artifact
boundary.

\paragraph{Adapter layer.}
An adapter is a thin export layer, not a retraining recipe. For a tokenizer
that exposes item-level codes, integration requires exporting one normalized
table and running preflight; the metric code is unchanged. The adapter records
method and dataset labels, source provenance, and one discrete token per
\sid level. Checkpoint-only releases require an upstream export step and do not
count as validated method coverage until item-level codes pass preflight.

\paragraph{Normalized contract.}
The required table has one stable item key and a consistent number of \sid
levels:
\begin{center}
\small
\texttt{item\_id}, \texttt{method}, \texttt{dataset}, \texttt{sid\_0}, \ldots,
\texttt{sid\_L} $\rightarrow$ \{\texttt{D1}, \ldots, \texttt{D5}\}.
\end{center}
Optional metadata and interactions use the same item keys and enable semantic,
collaborative-neighborhood, and head-tail slices. Paired mappings add D6;
generator-output or beam-trace tables add D7.
Current D1--D5 reports use a rectangular code table. Variable-length tokenizers
can be attached through padded or masked levels plus realized length. Without
generator traces, however, their serving interpretation remains limited to
structural D5 evidence.

\paragraph{Validation gate.}
Before metrics run, the validator checks item-count agreement, duplicate and
missing keys, discrete tokens, depth consistency, optional-table joins, and
source provenance. Public repositories may expose checkpoints, mappings,
codebooks, intermediate features, or partial releases; the gate prevents these
different objects from entering one comparison under the same label.

\paragraph{Probe engine and reports.}
A \emph{diagnostic probe} is a named metric family: D1 utilization, D2
aliasing, D3 neighborhood alignment, D4 popularity allocation, D5 structural
cost, D6 temporal churn, or D7 generation traces. A \emph{controlled mechanism
probe} is an input row designed to activate one known mechanism and check
whether a diagnostic reacts. Reports preserve the artifact label, dataset,
coverage, and provenance needed to compare valid rows. Table~\ref{tab:probe-actions}
states how D1--D5 support a next inspection action rather than a universal
pass/fail decision. D6 is implemented for paired refresh mappings; D7 is an
interface hook until generator outputs or beam traces are supplied.

\paragraph{Coverage and provenance.}
\tool reports tokenizer artifact rows, reference rows, and mechanism probes
under the same schema. LETTER and LC-Rec use validated official item-index
artifacts. The controlled GRID-style row uses RQ-KMeans. The bounded
ReSID/GAOQ row comes from its tokenizer stage and uses the same Musical items.
Reference and
control rows calibrate the probes; drift and non-Amazon rows exercise optional
interfaces. Literature-only methods remain future adapter targets.

Prior recommender resources such as RecList and Elliot motivate diagnostic and
reproducible evaluation beyond a single aggregate metric~\cite{chia2022reclist,anelli2021elliot}.
\tool differs by fixing the inspected object at a newer artifact boundary: the
item-to-code address space consumed by generative recommenders before a
reranker or generator is trained.

\section{Diagnostic Probes and Interpretation}

SID methods now optimize different properties of the exported address space.
Learned item indexing
and TIGER-style generative retrieval established the interface~\cite{hua2023indexids,rajput2023tiger}.
Later systems diversify how codes are learned and used through collaborative
or learnable tokenization~\cite{zheng2023lcrec,wang2024letter,liu2024elit},
behavior-semantic or multimodal evidence~\cite{wang2024eager,chen2026syngr}, and
collision or head-tail objectives~\cite{hu2026quasid,pan2026adasid,yan2026aktrec}.
Variable-length, drift, and deployment-facing designs add another set of
inspection needs~\cite{cheng2026capsid,baikalov2026staleness,ju2026snapchatsid}.
\tool organizes these concerns into reusable probes. The quantitative
diagnostic evidence appears in Tables~\ref{tab:musical-diagnostic}
and~\ref{tab:mechanism-probes}; Table~\ref{tab:probe-actions} turns each
mapping-level signal into a concrete inspection action.

\begin{table}[t]
\caption{Interpretation guide for D1--D5. Each probe identifies an artifact
signal and a follow-up check, not a universal pass/fail rule.}
\label{tab:probe-actions}
\scriptsize
\setlength{\tabcolsep}{2pt}
\begin{tabularx}{\columnwidth}{@{}>{\raggedright\arraybackslash}p{0.28\columnwidth}>{\raggedright\arraybackslash}p{0.32\columnwidth}>{\raggedright\arraybackslash}X@{}}
\toprule
Probe & Artifact signal & Next check \\
\midrule
\textbf{D1 Utilization} & Concentrated or dead code use & Per-level budgets,
collapse, and item coverage \\
\textbf{D2 Aliasing} & Shared full codes or broad prefixes & Alias groups by
popularity and co-occurrence \\
\textbf{D3 Neighborhood alignment} & Weak train-neighbor recovery & Same-dataset
controls, then candidate exposure \\
\textbf{D4 Popularity allocation} & Unequal head/mid/tail uniqueness & The
compressed bucket and intended allocation \\
\textbf{D5 Structural cost} & Large depth, fan-out, or active-prefix set &
Fixed-universe structure, then serving-stack latency \\
\bottomrule
\end{tabularx}
\end{table}

\paragraph{Notation.}
Let \(\mathcal{I}\) be the inspected item set and let a tokenizer export a code
sequence \(z(i)=(z_1(i),\ldots,z_L(i))\) for each item \(i\). We write the
level-\(\ell\) prefix as \(p_\ell(i)=(z_1(i),\ldots,z_\ell(i))\) and the full
alias set for code \(c\) as \(A(c)=\{i\in\mathcal{I}:z(i)=c\}\). The probes
below operate on \(z\), with metadata and interactions added only when a
diagnostic needs semantic or collaborative slices.

\paragraph{Mapping probes (D1--D5).}
D1, \emph{utilization}, reports per-level usage, prefix counts, imbalance
summaries, and dead or underused code indicators. D2, \emph{aliasing},
measures the fraction of items whose full \sid belongs to a non-singleton code
bucket, plus the same quantity for prefixes. This is an item-in-alias-set
profile rather than the number of duplicate code values; causal harm requires
intervention or downstream exposure checks beyond this mapping profile.
Because duplicate-code assignments are not uniformly harmful, the release also
includes a bounded interaction-qualified mechanism probe~\cite{hu2026quasid,zhang2026sidreliability}.

D3, \emph{neighborhood alignment}, asks whether prefix neighborhoods recover
train-only item co-occurrence neighbors. For each user, \tool forms item pairs
from bounded train interactions and counts their co-occurrence. It retains the
top-\(k\) directed neighbors per item, then reports the fraction whose neighbor
shares \(p_\ell(i)\). The weighted score averages edges, while the mean score
averages items. Metadata purity is auxiliary context, not collaborative
quality. D3 is a candidate-exposure diagnostic, not a ranking metric, and its
scale should be interpreted relative to controls on the same dataset.

D4, \emph{popularity allocation}, splits items by popularity and measures
whether full-code capacity and prefix structure are allocated differently across
head, mid, and tail items. D5, \emph{structural cost}, reports \sid length,
prefix fan-out, duplicate codes, and trie-like expansion pressure. These are
comparative structural signals, not a latency model. Actual serving cost
depends on how the generator consumes that structure; long, variable-length,
adaptive, and parallel SID designs therefore require implementation-specific
measurement~\cite{cheng2026capsid,zhang2026hgrec,wang2026sa2crq,hou2025parallel,hou2026expressiveness}.

\paragraph{Optional probes (D6--D7).}
D6, \emph{temporal churn}, computes mapping changes between tokenizer
refreshes for drift-aware settings~\cite{feng2026dact,baikalov2026staleness}.
It is implemented as an optional refresh-pair extension.
D7, \emph{generation traces}, is an interface designed to report invalid paths,
next-token entropy, and duplicate generated candidates. These reports require
generator-output tables or beam traces, so V0 provides the input hook but no
empirical D7 results.

\section{Resource Evaluation}

We evaluate \tool on two resource properties: diagnostic separation on a
fixed item universe and interface portability to released item-index
artifacts. The Musical protocol compares controlled GRID-style RQ-KMeans
exports and ReSID/GAOQ under one adapter contract, isolating capacity,
addressability, and neighborhood-alignment signals. LETTER and LC-Rec then exercise
the same contract on released item-index artifacts. Controls and mechanism
probes calibrate the structural, behavioral, and capacity-driven signals.

\paragraph{Same-item protocol.}
The case study uses the same 23,742 Musical items for the GRID-style and
ReSID/GAOQ rows.
GRID-style FT is a controlled feature-text export: ReSID processed feature text
is embedded and passed through a released GRID/RQ-KMeans-style residual k-means
module. We treat it as a controlled RQ-style artifact row rather than an
end-to-end GRID reproduction. ReSID is a bounded export from its FAMAE
tokenizer stage to GAOQ item codes on the same item universe. Holding the item
universe fixed isolates differences in the exported mappings.
The displayed GRID-style FT row is seed 42 with three width-64 levels
(64/64/64); its released source summary retains seeds 42--44. GRID-style
capacity uses the larger 32/1280/1280 per-level budget.
In Table~\ref{tab:musical-diagnostic}, D2 is full-code aliasing, D3 is level-1
weighted co-occurrence-prefix recall, D4 is tail unique-\sid ratio, and D5
lists active prefix counts by depth. Bold rows mark the primary same-item
artifact contrast; italics identify deterministic controls.

\begin{table}[t]
\caption{Artifact profiles on the same 23,742 Musical items. Bold rows form the
primary contrast; italics mark controls. D2 is full-code aliasing ($\downarrow$),
D3 is level-1 neighborhood alignment ($\uparrow$), D4 is tail uniqueness
($\uparrow$), and D5 gives active-prefix counts by depth.}
\label{tab:musical-diagnostic}
\scriptsize
\setlength{\tabcolsep}{1.8pt}
\begin{tabular*}{\columnwidth}{@{\extracolsep{\fill}}lrrrrl@{}}
\toprule
Artifact profile & Unique SIDs & D2 alias. & D3 align. & D4 tail & D5 prefix counts \\
\midrule
\multicolumn{6}{@{}l}{\emph{RQ-style artifacts}} \\
\textbf{GRID-style FT} & 3,749 & 0.977 & 0.055 & 0.370 & 64/3.4k/3.7k \\
GRID-style capacity & 9,874 & 0.779 & 0.080 & 0.639 & 32/9.3k/9.9k \\
RQ-min reference & 17,247 & 0.440 & 0.065 & 0.883 & 32/2.4k/17.2k \\
\addlinespace[1pt]
\multicolumn{6}{@{}l}{\emph{ReSID/GAOQ artifact}} \\
\textbf{ReSID/GAOQ} & 23,742 & 0.000 & 0.154 & 1.000 & 32/1.3k/23.7k \\
\addlinespace[1pt]
\multicolumn{6}{@{}l}{\emph{Calibration controls}} \\
\emph{Category-prefix} & 23,742 & 0.000 & 0.447 & 1.000 & 30/83/313/23.7k \\
\emph{Popularity-balanced} & 22,707 & 0.086 & 0.303 & 0.962 & 4/1.0k/22.7k/22.7k \\
\emph{Hash-collision} & 256 & 1.000 & 0.004 & 0.032 & 256/256/256/256 \\
\bottomrule
\end{tabular*}
\end{table}

\begin{table}[t]
\caption{Mechanism checks for D2, D1/D4, and D5. Each arrow compares paired
conditions within one control, not tokenizer methods.}
\label{tab:mechanism-probes}
\scriptsize
\setlength{\tabcolsep}{2pt}
\begin{tabularx}{\columnwidth}{@{}>{\raggedright\arraybackslash}p{0.31\columnwidth}>{\centering\arraybackslash}p{0.13\columnwidth}>{\raggedright\arraybackslash}X@{}}
\toprule
Mechanism control & Probe & Observed response \\
\midrule
Interaction-qualified aliasing & D2 & shared-user lift: hash 1.19x $\rightarrow$ co-occurrence 3.86x \\
Head-reserved capacity & D1/D4 & head unique 1.000 $\rightarrow$ tail unique 0.028 \\
Variable depth & D5 & max-depth 12,010 $\rightarrow$ realized 7,914 \\
\bottomrule
\end{tabularx}
\end{table}

Table~\ref{tab:musical-diagnostic} makes the same-item rows comparable under
one schema. The GRID-style and ReSID/GAOQ rows form a controlled tokenizer-artifact
contrast; the capacity ablation, RQ-min reference, and controls explain which
parts of the contrast come from capacity, addressability, or neighborhood structure.
In this setup, GRID-style FT exposes high aliasing pressure and limited tail
capacity, while the ReSID/GAOQ row exports one full code per item and has
higher D3-L1 collaborative prefix recovery. ReSID/GAOQ and the category-prefix
control show the zero-aliasing endpoint: each item has a distinct full code,
but that property alone does not establish recommendation quality.

The GRID-style capacity row isolates the budget component of the original
GRID-style pressure. With a larger 32/1280/1280 budget, unique full codes rise to
9,874 and D2 aliasing drops to 0.779. The export still provides only 9,874
unique codes for 23,742 items, so capacity explains only part of the pressure.
Because full codes remain shared, this is an artifact-profile comparison rather
than a method-quality claim. RQ-min is a minimal residual-quantization reference
adapter that passes the same validator;
it tests adapter independence on the full Musical item universe and is reported
as a reference row rather than an additional tokenizer line.

\paragraph{Findings.}
Addressability and behavioral neighborhood alignment are distinct artifact
properties. ReSID is structurally item-unique, and the
category-prefix control is also alias-free, but D3 separates their profiles.
ReSID is a learned export, whereas the category row is a deterministic control
for interpreting the metric scale. The non-learned category-prefix row recovers
more level-1 co-occurrence
neighbors (0.447) than ReSID (0.154) or either GRID row (0.055--0.080). The
contrast reveals a neighborhood-alignment gap: learned or item-unique addresses do
not necessarily group behaviorally related items. This distinction becomes
visible when the mapping is inspected directly.

The RQ-style rows also separate capacity from alignment.
D2 aliasing falls from 0.977 (GRID-style FT) to 0.779 (GRID-style capacity) and
0.440 (RQ-min reference), while D3 stays in the narrow 0.055--0.080 range. Capacity can
increase and aliasing can fall without producing behaviorally aligned prefix neighborhoods,
making D2 and D3 complementary diagnostics. The ablation separates a capacity
interpretation from the remaining artifact-profile gap.
Table~\ref{tab:mechanism-probes} shows the expected responses directly: the
interaction-qualified collision lift rises from 1.19x for hash collisions to
3.86x for the co-occurrence row; head-reserved capacity preserves head
uniqueness while compressing the tail; and realized depth reduces active
prefixes relative to maximum-depth accounting.

Released item-index adapters test the resource beyond the same-item case.
LETTER and LC-Rec Instruments artifacts both pass the same validation path on
9,922 items; both expose 9,897 unique full \sid{}s, with D3-L1 0.109 for LETTER
and 0.052 for LC-Rec. They appear in the catalog rather than the same-item
table because they use a different item universe; their role is to show that
\tool ingests released tokenizer mappings as well as locally generated
contrasts.

D3 scale is dataset-dependent. On an
All\_Beauty 20k vertical, a coarse category-prefix control reaches D3-L1
0.968, while GRID/RQ-KMeans feature-text rows reach 0.081, 0.087, and 0.090
across seeds 42--44. Compared with Musical, this result shows how strongly the
available category signal can shape D3.

Bounded fixed-reranker probes on 5,000 users then connect D3 to candidate
exposure. Across artifact and control rows, D3 correlates with train-only
prefix candidate recall at depth 1 (Spearman 0.976 across eight Musical rows
and 1.000 across six All\_Beauty rows). For the fixed reranker at \(K=20\),
the rank correlations are 0.970/0.952 for Recall/NDCG on Musical and
0.657/0.657 on All\_Beauty. These small-row correlations are diagnostic checks,
not population estimates; they connect D3 to exposure before final ranking.
For a weak D3 row, the next check is a same-dataset control and, when useful,
the fixed-reranker exposure probe before committing to generator training. It
does not by itself reject a tokenizer or predict trained-generator quality.
A Sports 20k GRID export adds a cross-domain check with D3-L1 0.055, an
\sid{} duplicate rate of 0.592, and D5 prefix counts of 128/7986/8165. The
reproducibility matrix
separately indexes MovieLens schema and DACT churn checks.

\section{Availability and Limitations}

The public repository at \url{https://github.com/jdding/sidinspector} provides
the MIT-licensed v1.0 package, adapter template, examples, tests, documentation,
and clean-checkout verifier. Users can validate new item-to-\sid{} mappings and
produce D1--D5 reports without local caches. Released summaries and run manifests
feed Tables~\ref{tab:musical-diagnostic} and~\ref{tab:mechanism-probes}; one
command rebuilds both table CSVs, and the verifier checks every row. These
snapshots reproduce the tables, not upstream tokenizer training or omitted
caches.

The resource supports three uses. A downstream researcher can flag missing
mappings, inconsistent depth, alias groups, weak collaborative-prefix recovery,
or tail compression before generator training. A method author can emit the
normalized SID table, attach optional metadata or interactions, and reuse the
validator and D1--D5 engine. A reader can run the quickstart, table builder, and
package verifier from a clean checkout. Table~\ref{tab:probe-actions} maps each
signal to a next inspection action; the toolkit supplies no universal threshold.

The four tokenizer lines have explicit provenance: LETTER and LC-Rec use
validated official item-index artifacts, GRID-style is a controlled RQ--KMeans
export, and ReSID/GAOQ is a bounded tokenizer-stage export under the same-item
Musical protocol. RQ-min and the mechanism rows are controls, not extra
named-method coverage. All\_Beauty, Sports, MovieLens, and DACT check dataset
effects, schema portability, and optional churn, not method ranking.

The examples separate addressability, neighborhood alignment, popularity
allocation, and structural pressure at the artifact level. Their scope remains
diagnostic: D2 is not causal collision-harm evidence; D3 and the fixed-reranker
probe measure candidate exposure rather than trained-generator quality; D5 is
not serving latency; and D7 remains an input hook. V0 therefore delivers a
versioned mapping-first contract, validation gate, actionable D1--D5 reports,
and reproducible evidence. Downstream interventions, broader artifact coverage,
generator traces, and decoding-stack evaluation remain future work.

\section*{GenAI Usage Disclosure}
Generative AI tools assisted with language editing and code debugging. All
experimental results were produced by author-written scripts and verified
against saved artifacts.

\begingroup
\sloppy
\hbadness=10000
\bibliographystyle{ACM-Reference-Format}
\bibliography{references}
\endgroup

\end{document}